\documentstyle[prl,aps,epsf,floats]{revtex}
\begin{document}
\draft
\twocolumn[\hsize\textwidth\columnwidth\hsize\csname @twocolumnfalse\endcsname
\title{Real-time correlation functions from imaginary-time evolution} 

\author{P.\,E.\,Kornilovitch}
\address{
Hewlett-Packard Laboratories, MS 1L-12, 1501 Page Mill Road, Palo Alto, 
California, 94304
}

\date{\today}
\maketitle
\begin{abstract}

The problem of calculating real-time correlation functions is 
formulated in terms of an imaginary-time partial differential 
equation.  The latter is solved analytically for the perturbed 
harmonic oscillator and compared with the known exact result.  
The first order approximation for the short-time propagator is 
derived and used for numerical solution of the equation by a 
Monte Carlo integration.  In general, the method provides a 
reformulation of the dynamic sign problem, and is applicable 
to any two-time correlation function including single-particle, 
density-density, current-current, spin-spin, and others.  The 
prospects of extending the technique onto multi-dimensional 
problems are discussed.  

\end{abstract}
\vskip2pc]
\narrowtext

\section{Introduction}
\label{sec:one}

Quantum Monte Carlo (QMC) simulations proved very successful in 
studying many-body systems with realistic inter-particle potentials.  
Specific QMC methods have been developed for interacting electrons 
(for a recent review see \cite{Foulkes}), liquid $^4$He \cite{Ceperley}, 
alkali Bose-condensates in harmonic traps \cite{Krauth}, electron-proton 
plasma \cite{Magro} and other systems.  These methods yield accurate 
values of various physical properties, such as the cohesive energies 
of solids \cite{Leung} or effective masses of defects 
\cite{Bonisegni}, which are in excellent agreement with experiment.  
In model problems including spin \cite{Evertz}, strong correlation 
\cite{Blankenbecler}, and electron-phonon \cite{Kornilovitch,Prokofev} 
models, QMC usually handles much larger systems than other methods, 
treats equally well simple and complex interactions, and produces 
accurate results.  
 
A major difficulty faced by QMC is its inability to compute reliably 
dynamic properties of quantum mechanical systems.  This is because 
QMC operates in {\em imaginary} time.  For those quantities that do 
not directly translate in an imaginary-time language, the simulation 
noise quickly exceeds the signal rendering the calculation impossible.  
This difficulty is known as the ``dynamical sign problem''.  It is 
usually dealt with by computing correlation functions 
in imaginary time first, and then continuing the results into the 
real-time domain.  For the latter step, Pade approximants \cite{Vidberg}, 
the maximum entropy \cite{Jarrell}, singular-value decomposition 
\cite{Creffield}, and stochastic methods \cite{Mishchenko} are employed.  
Unfortunately, the analytic continuation of noisy data is an ill-posed 
mathematical procedure that does not lead to a unique solution.  
Sometimes such an uncertainty results in spectacular failures of the 
reconstruction strategy.  For instance, even the most advanced and 
popular method, the maximum entropy, cannot resolve the two peaks of 
the dynamic structure factor in liquid helium \cite{Ceperley}.  This 
is a consequence of dealing with the mathematically ill-defined problem.  
Another difficulty is assigning meaningful error bars to the continuation
procedure.    

Thus the reconstruction procedure, while sometimes working reasonably
well (especially in model systems), remain approximate, spoiling 
the rigor of QMC.  What would improve the situation is the direct link
between simulations and dynamic properties or rendering the problem
mathematically well-posed.  One such method was proposed by Mak and Egger
who introduced ``the multilevel blocking algorithm'' to deal with the 
dynamic sign problem \cite{Mak}. (A similar method was proposed for 
the fermion sign problem \cite{Egger}.)  Essentially, this was an 
attempt of straightforward evaluation of real-time rapidly oscillating
path integrals that express the time dependent transition amplitudes.
This method requires very large computer memory even for simple
one-particle systems.  Since the sign problem becomes more severe 
with increasing the size of the system, the blocking algorithm may
not sustain the increase in the number of simulated particles.  It
leaves the stimulus for the search of other alternatives.     
  
In this paper, I propose another approach to alleviating the dynamic 
sign problem.  The calculation of real-time correlators is formulated
as a mathematically well-defined initial value problem for the operator 
$(\omega-H)^2$ which is of the fourth order in spatial coordinates.  
The solution of the corresponding {\em imaginary}-time evolution equation 
yields directly the frequency-dependent spectral function, thereby 
eliminating the need for the time Fourier transformation, analytic 
continuation and other intermediate procedures.  The basic idea is 
very general and is applicable to any quantum mechanical system and 
any two-time correlation function.  It is described in 
Section~\ref{sec:two}.  In Section~\ref{sec:three} I apply the method 
to an exactly solvable case of disturbed one-dimensional harmonic 
oscillator.  Solving the new equation in more general cases is hard 
and require stochastic Monte Carlo methods.  The arising difficulties 
and their possible solutions are discussed in 
Sections~\ref{sec:four}-\ref{sec:seven}.

\section{A differential equation for the spectral function}
\label{sec:two}

Consider a typical problem of many-body physics, that of finding 
the single-particle zero-temperature correlator in an interacting 
$N$-particle system:
\begin{eqnarray}
K({\bf q},t) & = & \langle G \vert \, c_{\bf q}(t) c^{\dagger}_{\bf q}(0) 
\vert G \rangle       \nonumber \\
& = & e^{iE_Gt} \langle G \vert \, c_{\bf q} e^{-iHt} c^{\dagger}_{\bf q} 
\vert G \rangle .
\label{one}
\end{eqnarray}
Here $|G\rangle$, $E_G$, and $H$ are the ground state, its energy, 
and the full Hamiltonian of the system, respectively.  The particles 
could be either fermions or bosons.  The operator $c^{\dagger}_{\bf q}$ 
creates an extra particle with momentum {\bf q}, and operator $c_{\bf q}$ 
destroys it.  [It has to be emphasized that the single-particle momentum 
operators are chosen for definiteness only.  In fact, everything that 
will be said and written about those operators will be equally valid 
for any pair of operators.]  The notation $\hbar=1$ is used throughout 
the paper.  $K({\bf q},t)$ describes the decay of 
$|\Psi\rangle = c^{\dagger}_{\bf q} \vert G \rangle$ with time due to 
$|\Psi\rangle$ not being an eigenstate of $H$.  An equivalent amount of 
information is contained in the spectral density function 
$A({\bf q},\omega)$ which is a Fourier transformation of $K({\bf q},t)$:
\begin{eqnarray}
A({\bf q},\omega) & = & 
\int^{\infty}_{-\infty} K({\bf q},t) e^{i\omega t} \, dt \nonumber \\
& = & 
(2\pi) \langle G\vert c_{\bf q}\, \delta(\omega + E_G - H) \, 
c^{\dagger}_{\bf q} \vert G\rangle .
\label{two}
\end{eqnarray}
Compare the structures of Eqs.~(\ref{one}) and (\ref{two}).  The 
correlation function is complex and has an oscillating kernel $e^{-iHt}$.  
The latter is the origin of the sign problem since big positive and 
negative contributions tend to cancel each other with a very small 
residue.  On the contrary, $A({\bf q},\omega)$ is positive definite 
and has a non-oscillating kernel, a delta-function, which intuitively 
makes it a better computational object.  There is another, physical 
reason to prefer the function $A({\bf q},\omega)$ over $K({\bf q},t)$.  
The time-dependent properties themselves are rarely measured 
experimentally.  What is usually inferred from experiments are namely 
the energy-dependent spectral functions, or related to them response 
functions.  Thus rather than compute $K$s with subsequent numerical
Fourier transformations, it would be much better to be able to compute
$A$s directly.  Besides, the main interest represent low-energy responses 
at small $\omega$s.  This requires the knowledge of $K$s at {\em large} 
times $t$ that is precisely where the sign problem is worst. 

The above considerations suggest to compute $A({\bf q},\omega)$ instead 
of $K({\bf q},t)$ \cite{Matthew}.  However, while being formally correct, 
Eq.~(\ref{two}) is impractical due to the presence of the operator function 
$\delta(\omega+E_G-H)$.  There is no way of replacing $H$ with an explicit 
functional of the particle coordinates (which is required for numerical 
integration) other than using a different representation of the 
delta-function that would allow evaluation of the matrix elements of $H$ 
in various powers.  The common Fourier expansion of the delta-function 
would take the situation back to Eq.~(\ref{one}) and to the sign-problem.  
However, it is not necessary to use an integral representation.  Instead, 
I use a {\em limit} representation of the delta-function, specifically
the limit of a gaussian.  After that Eq.~(\ref{two}) becomes
\begin{equation}
A({\bf q},\omega) = 
(2\pi) \lim_{\alpha \rightarrow \infty}  
\langle G\vert c_{\bf q} \, \sqrt{\frac{\alpha}{\pi}} 
e^{-\alpha (\omega+E_G-H)^2} \, c^{\dagger}_{\bf q}\vert G\rangle .
\label{four}
\end{equation}
Taking the limit now amounts to solving the {\em imaginary}-time 
evolution equation
\begin{equation}
- \frac{\partial \psi}{\partial \alpha} = (\omega+E_G-H)^2 \psi ,
\label{five}
\end{equation}
supplemented with the initial condition $\psi(\alpha=0) = 
c^{\dagger}_{\bf q}\vert G\rangle$. 

The equation (\ref{five}) and representation (\ref{four}) are the main 
results of the paper.  They show that dynamic correlators can in 
principle be obtained by solving an imaginary-time partial differential
equation.  The equation is supplemented by an initial condition and
therefore it possesses a unique solution.  The whole problem is thus
mathematically well-defined.  The full computational procedure is as 
follows: 
(i) Obtain a ground state wave function $|G\rangle$ and energy $E_G$
with whatever equilibrium method that works for the given system. 
(ii) Construct new state 
$|\Psi\rangle = c^{\dagger}_{\bf q}\vert G\rangle $. 
(iii) Solve Eq.~(\ref{five}) using $|\Psi\rangle$ as an initial
condition.
(iv) Perform the $\alpha \rightarrow \infty$ limit and compute the
scalar product with $\langle \Psi |$ to get the spectral density for
given frequency $\omega$.
(v) Compute the response functions via fluctuation-dissipation and
Kramers-Kronig relations.

Let me discuss the general properties of Eq.~(\ref{five}).  First of 
all, this equation is of the first order in ``time'' variable $\alpha$ 
and {\em fourth} order in spatial coordinates because of the term $H^2$.  
Mathematically, this is a parabolic equation ``correct in Petrovski sence'' 
\cite{Fedoriuk}.  Its properties are quite different from the familiar 
second order parabolic (diffusion) equations.  In particular, its Green
function is no positive-definite but instead can have positive and negative 
regions.  Because of that, the proposed method does not solve the sign
problem.  Rather, it transforms the dynamic, or ``time'', sign problem
into a ``space'' sign problem which resembles the fermion one.  More on 
this in Section~\ref{sec:four}.  It is instructive to compare 
equation (\ref{five}) with the imaginary time Schr\"odinger equation: 
$- \partial \psi/\partial \beta = (H-E_G) \psi$. The latter's formal
solution is $\psi(\beta) = e^{-\beta(H-E_G)} \psi(0)$.  It shows that
in the $\beta \rightarrow \infty$ limit, only the ground state survives
because all the other states are exponentially suppressed.  This well
known fact is employed for instance in Projected Monte Carlo to extract
the ground state out of an arbitrary initial state $\psi(0)$.  The 
crucial observation here is that the energies of other states are all 
greater than $E_G$ which makes the exponent of the evolution operator 
be negative for all but the ground state.  But what if one wants
to retain not the ground state but a state (or states) with energy
$(E_G+\omega)$?  In this case one would need a similar evolution operator
with the exponent being a negative and even function of $(E_G+\omega-H)$.
The simplest function is quadratic one which results in precisely the 
exponential kernel of Eq.~(\ref{four}).  Thus the physical meaning of the
representation (\ref{four}) is that in the course of evolution only
excited states with excitation energy $\omega$ survive and all the other
states with excitation energies {\em above and below} $\omega$ are
exponentially suppressed.  Notice also that while the variable $\beta$
in the Schr\"odeinger equation is related to the inverse temperature,
the variable $\alpha$ in Eq.~(\ref{five}) has no clear physical meaning.
The dimensionality of $\alpha$ is (energy)$^{-2}$. 
Lastly, one may think that many other variations of Eqs.~(\ref{four}) 
and (\ref{five}) are possible since there are infinitely many limit 
representation of the delta-function.  This does not seem to be the 
case.  All other common representation of $\delta(x)$ involve either 
non-analytical or non-polynomial functions (examples are $e^{-\alpha |x|}$ 
and $\alpha/(1+x^2\alpha^2)$.  Both cases make it difficult to interpret
the resulting formulae in terms of a simple evolution equation.  
Comparable or even better representations may exist but I could not
find any.  
    
I now rederive Eq.~(\ref{four}) differently, by demonstrating
its equivalence to another definition of the spectral function.
Let $|\phi_m\rangle$ be a set of eigenstates for the $(N+1)$-particle 
system, $H |\phi_m\rangle = E_m |\phi_m\rangle$.  In general, 
$|\Psi\rangle$ is not an eigenstate of $H$ but it is always 
expandable in $|\phi_m\rangle$:
\begin{equation}
|\Psi\rangle = \sum_m a_m |\phi_m\rangle = \sum_m 
|\phi_m\rangle \langle \phi_m | c^{\dagger}_{\bf q} | G \rangle.
\label{six}
\end{equation}
Upon substitution of the expansion in Eq.~(\ref{four}) the
Hamiltonian in the exponent is replaced with $E_m$.  After that
the $\alpha \rightarrow \infty$ limit results in a delta-function
$\delta(\omega+E_G-E_m)$ and Eq.~(\ref{four}) becomes
\begin{equation}
A({\bf q},\omega) = (2\pi) \! \sum_m \langle G | c_{\bf q} | 
\phi_m\rangle \langle \phi_m | c^{\dagger}_{\bf q} | G \rangle \, 
\delta(\omega + E_G - E_m) ,
\label{seven}
\end{equation}
which is the usual definition of $A({\bf q},\omega)$ \cite{Mahan}. 

The formalism presented above is valid at zero temperature only.
Its finite-temperature generalization is not straightforward
and has to be addressed separately.  The correlation function is
defined as
\begin{equation}
K({\bf q},t) = \frac{1}{Z} \sum_n 
\langle n \vert e^{iHt} c_{\bf q} e^{-iHt} c^{\dagger}_{\bf q} 
e^{-\beta H} \vert n \rangle ,
\label{sevenone}
\end{equation}
where $|n\rangle$ is a complete set of the $N$-particle system,
$\beta$ inverse temperature, and $Z=\sum_n \exp{(-\beta E_n)}$
the partition function.  A new difficulty in comparison with 
Eq.~(\ref{one}) is the second real-time operator $e^{iHt}$ which acts 
in a different state space than $e^{-iHt}$.  Therefore, a time 
Fourier transform of Eq.~(\ref{sevenone}) does not lead to a convenient
representation of the spectral density similar to Eq.~(\ref{two}).
Put differently, $e^{-iHt}$ cannot be replaced with a $c$-number
because it now acts not on a single state but on the whole set 
$\langle n|$ which is not supposed to be known in its entirety.      
Still, one can proceed further by formally separating the time
variables in the two operators.  Introducing an auxiliary 
delta-function $1 = \int dt' \delta(t-t')$, replacing 
$t \rightarrow t'$ in $e^{-iHt}$, using $\delta(t-t') = 
\int \frac{d\varepsilon}{2\pi} e^{-i\varepsilon (t-t')}$, 
and integrating over $t'$ one obtains
\begin{eqnarray}
K (&{\bf q},t&) =       \nonumber \\ 
& \displaystyle \frac{1}{Z} & \int^{\infty}_{-\infty} \!\!\! d\varepsilon 
\, e^{-i\varepsilon t} \sum_n \langle n \vert e^{iHt} c_{\bf q} 
\delta(\varepsilon-H)  c^{\dagger}_{\bf q} e^{-\beta H} \vert n \rangle .
\label{seventwo}
\end{eqnarray}
A Fourier transformation yields for the the spectral density
\begin{eqnarray}
 A (&{\bf q},\omega&) = \nonumber \\
& \displaystyle \frac{2\pi}{Z} & \int^{\infty}_{-\infty} \! d\varepsilon 
\sum_n \langle n \vert \, \delta(\omega-\varepsilon +H) \, c_{\bf q} 
\, \delta(\varepsilon-H) \, c^{\dagger}_{\bf q} e^{-\beta H} 
\vert n \rangle           \nonumber \\
 = & \displaystyle \frac{2\pi}{Z} & \lim_{\alpha_1 \rightarrow \infty}
\lim_{\alpha_2 \rightarrow \infty}
\sqrt{\frac{\alpha_1}{\pi}} \sqrt{\frac{\alpha_2}{\pi}}
\int^{\infty}_{-\infty} \! d\varepsilon \sum_n  \nonumber \\
& \times & \langle n \vert \, e^{-\alpha_1(\omega-\varepsilon +H)^2} 
\, c_{\bf q} \, e^{-\alpha_2(\varepsilon-H)^2} \, c^{\dagger}_{\bf q} 
e^{-\beta H} \vert n \rangle  .
\label{seventhree}
\end{eqnarray}
There are {\em three} imaginary-time evolution processes in the last 
equation.  First one, starting with an arbitrary state $|n\rangle$,
during time $\beta$, and under operator $H$.  Second one, starting
from the finite state of the first process plus an extra
particle with momentum {\bf q}, during infinite time, and
under operator $(\varepsilon-H)^2$.  Third one, starting with the
final state of the second process minus a particle with momentum
{\bf q}, during infinite time, and under operator 
$(\omega-\varepsilon +H)^2$.  In the end, the scalar product with 
$\langle n|$ and integration over the real variable $\varepsilon$ 
have to be done.  In the zero-temperature limit, the operator 
$e^{-\beta H}$ projects out all but the ground states, 
$e^{-\beta H} |n\rangle = e^{-\beta E_G} |G\rangle \delta_{nG}$,
and then the factor $e^{-\beta E_G}$ cancels $Z$.  After that,
the leftmost $H$ in Eq.~(\ref{seventhree}) is replaced with
$E_G$ and the limit in $\alpha_1$ and integration over 
$\varepsilon$ are easily performed, thereby reducing 
Eq.~(\ref{seventhree}) to the previous expression (\ref{four}).

The finite-temperature generalization is achieved at the expense of 
the second auxiliary evolution and an extra integration.
This makes any practical realization of the method much more
difficult than at zero temperature.  Nevertheless,  
Eq.~(\ref{seventhree}) is exact and contains only real 
exponentials. 

In conclusion of this section, let me reiterate that nowhere the 
specifics of the single particle operators (fermionic or bosonic) have 
been used.  {\em Everything} that has been said about $c_{\bf q}$ and 
$c^{\dagger}_{\bf q}$ is equally applicable to {\em any} pair of 
operators $\hat O_1$ and $\hat O_2$.  A generic expression for the 
zero-temperature spectral density is a straightforward 
generalization of Eq.~(\ref{four}):
\begin{equation}
A_{\hat O_1 \hat O_2}(\omega) = 
(2\pi) \lim_{\alpha \rightarrow \infty}  
\langle G \vert \hat O_2 \, \sqrt{\frac{\alpha}{\pi}} 
e^{-\alpha (\omega +E_G - H)^2} \, \hat O_1 \vert G\rangle .
\label{fourteen}
\end{equation}
There are no restrictions on the choice of the operators $\hat O_1$ and 
$\hat O_2$.  They could be single-particle, two-particle, spin, or other. 
In the case of the density or current operators, Eq.~(\ref{fourteen}) 
opens the exciting possibility to compute dynamic structure factors and 
conductivities from first principles.

\section{Perturbed harmonic oscillator: Analytical solution}
\label{sec:three}

The purpose of this Section is to solve a non-trivial example to 
demonstrate Eqs.~(\ref{four}) and (\ref{five}) at work.  Consider
a harmonic oscillator with frequency $\Omega$ and mass $M$.  
It can be either free of subjected to a constant force of magnitude
$g\sqrt{2M\hbar\Omega^3}$, depending on whether an additional particle 
is present in the vicinity of the oscillator or not.  The Hamiltonian 
reads
\begin{equation}
H_x = \Omega \left( - \frac{1}{2} \frac{\partial^2}{\partial x^2} 
+ \frac{1}{2} x^2 - \sqrt{2} g x \, c^{\dagger} c \right) ,
\label{eight}
\end{equation}
where the coordinate $x$ is measured in units of $\sqrt{\hbar/(M\Omega)}$
and $g$ is the dimensionless coupling constant.  This Hamiltonian is 
known from the independent boson model \cite{Mahan}.  It can also
be thought of as the on-site Holstein polaron \cite{Holstein}.   
Let the oscillator be in its ground state and the particle absent at
$t<0$.  The state of the system is $|G\rangle = |0_{\rm part},0_x\rangle 
= |0_{\rm part}, \frac{1}{\pi^{1/4}} e^{-x^2/2} \rangle$ with
energy $E_G = \frac{1}{2}\Omega$.  At time 0 the particle is placed
on the oscillator creating state 
$|\psi(x,\alpha=0)\rangle = |1_{\rm part}, 0_x \rangle$.  At this time
the oscillator begins to experience the force.  At time $t$ the particle 
is removed and the oscillator is left free but in one of the excited
states $|m_x\rangle$.  So between 0 and $t$ the Hamiltonian is
\begin{equation}
H_x = \Omega \left( - \frac{1}{2} \frac{\partial^2}{\partial x^2} 
+ \frac{1}{2} x^2 - \sqrt{2} g x \right) .
\label{eight_one}
\end{equation}
Our objective is to compute the spectral function $A(\omega)$ of
operators $c{\dagger}$ and $c$.  According to the general scheme,
one requires to solve equation (\ref{five}) with $H_x$ from 
(\ref{eight_one}) and initial condition $|0_x\rangle$.  This can be 
done by introducing first a new variable $y=x-\sqrt{2}g$ and then 
bosonic operators $b_y$ and $b^{\dagger}_y$.  The transformed 
Hamiltonian is $H_y = \Omega (b^{\dagger}_y b_y - g^2 + \frac{1}{2})$.  
After expansion of the solution in the eigenstates of $b^{\dagger}_y b_y$, 
$|\psi\rangle = \sum^{\infty}_{m=0} a_m(\alpha) |m_y\rangle$, 
Eq.~(\ref{five}) becomes:
\begin{eqnarray}
- \sum^{\infty}_{m=0} \frac{\partial a_m(\alpha)}{\partial \alpha} 
|m_y\rangle 
& = & \!\! \sum^{\infty}_{m=0} a_m(\alpha) 
      (\omega + g^2\Omega -\Omega b^{\dagger}_y b_y )^2 |m_y\rangle 
      \nonumber \\ 
  = \sum^{\infty}_{m=0} &a_m&(\alpha) 
    (\omega + g^2\Omega -\Omega m )^2 |m_y\rangle . 
\label{nine}
\end{eqnarray}
Equations with different $m$ are separated and easily integrated
yielding
\begin{eqnarray}
\psi(x,\alpha) & = & \sum^{\infty}_{m=0} a_m(0) 
e^{-\alpha (\omega + g^2\Omega -\Omega m )^2}  \nonumber \\ 
& \times & \frac{1}{\pi^{1/4}\sqrt{2^m m!}} H_m(x-\sqrt{2}g) 
e^{-\frac{1}{2}(x-\sqrt{2}g)^2} ,
\label{ten}
\end{eqnarray}
where $H_m(x)$ are the Hermite polynomials.
The coefficients $a_m(0)$ are found from the initial condition
$\psi(x,0) = $$\pi^{-1/4} \exp{(-x^2/2)}$: 
\begin{equation}
a_m(0) = \langle m_y | 0_x \rangle = 
\frac{(-\sqrt{2}g)^m}{\sqrt{2^m m!}} e^{-\frac{1}{2}g^2} .
\label{eleven}
\end{equation}
\begin{figure}[t]
\begin{center}
\leavevmode
\hbox{
\epsfxsize=8.4cm
\epsffile{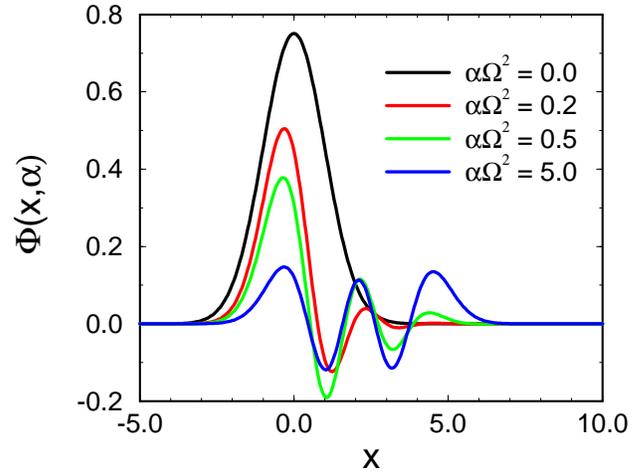}
}
\end{center}
\vspace{-0.5cm}
\caption{
The wave function (\ref{ten})-(\ref{eleven}) at different ``times''
$\alpha$.  $g=1.5$ and $\omega=2.0\,\Omega$.  Note how negative pockets 
develop with time.  After $\alpha=5.0$ the solution changes only
slightly. 
}
\label{fig1}
\end{figure}
\begin{figure}[t]
\begin{center}
\leavevmode
\hbox{
\epsfxsize=8.4cm
\epsffile{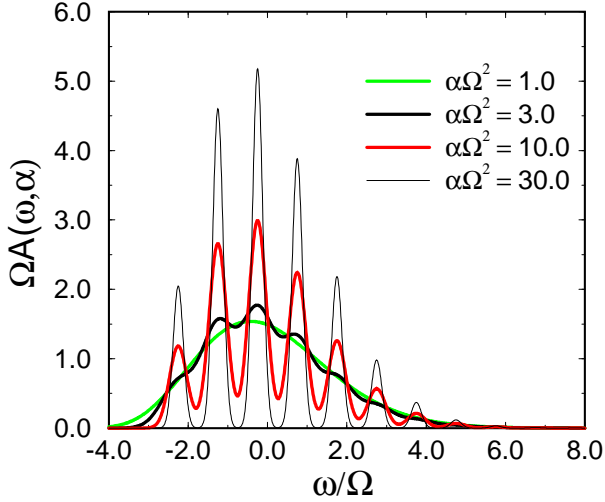}
}
\end{center}
\vspace{-0.5cm}
\caption{
The spectral function (\ref{twelve}) of the perturbed harmonic
oscillator for different values of the parameter $\alpha$.  $g=1.5$.  
The first (leftmost) peak appears at $\omega=-g^2\Omega$.  The area 
under all the curves is $2\pi$.  This is how the correct result 
would emerge in the course of numerical solution of Eqs.~(\ref{four}) 
and (\ref{five}). 
}
\label{fig2}
\end{figure}

Figure~\ref{fig1} shows the solution (\ref{ten})-(\ref{eleven}) as
a function of ``time'' $\alpha$.  Note the appearance of negative 
regions.  Finally, the scalar product 
$\langle 0_x| \psi(x,\alpha) \rangle$ yields the 
``$\alpha$-dependent spectral density''
\begin{equation}
A(\omega,\alpha) = (2\pi) e^{-g^2} \sum^{\infty}_{m=0} \frac{g^{2m}}{m!}
\sqrt{\frac{\alpha}{\pi}} e^{-\alpha (\omega+g^2\Omega-\Omega m)^2} . 
\label{twelve}
\end{equation}
Figure \ref{fig2} illustrates the changes in $A(\omega,\alpha)$ as 
$\alpha$ progresses from 0 to $\infty$.  The solution evolves 
from the identical zero through a smooth single peak to a highly 
non-analytical grid of delta-functions.  The 
$\alpha \rightarrow \infty$ limit of Eq.~(\ref{twelve}) results in
\begin{equation}
A(\omega) = (2\pi) e^{-g^2} \sum^{\infty}_{m=0} \frac{g^{2m}}{m!}
\delta(\omega+g^2\Omega-\Omega m) , 
\label{thirteen}
\end{equation}
which is known to be the correct spectral density \cite{Mahan} 
(chapter 4).  

Two comments are appropriate at this point.  The first one concerns 
the solvability of the equation (\ref{five}).  In the above example,
the equation was solvable because the oscillator before, during, and
after the perturbation was an integrable system which eigenfunctions
were well known.  Surely, for {\em any} such a system 
where all the eigenfunctions are known, the equation (\ref{five})
would be solvable as easily as for the oscillator, and the results
would be expressible in the form of a spectral expansion similar
to Eqs.~(\ref{ten})-(\ref{twelve}).  The second observation is 
about the differences between systems with discrete and continuous 
spectrum.  In the case of discrete spectrum, as in the above example, 
the spectral function must have the form of a grid of delta-functions,
see Eq.~(\ref{seven}).  This is realized in the present formalism
in the following way.  For certain frequencies $\omega$, the solution
of equation (\ref{five}) has a finite overlap with the final state
$\langle G|c_{\bf q}$ {\em even} in the limit $\alpha \rightarrow \infty$.
Then the factor $\sqrt{\alpha}$ results in infinite limit values of
the spectral function for these resonant frequencies.  For systems
with continuous spectra, including all real systems, the overlap and
consequently the solution of (\ref{five}) {\em must decay as} 
$1/\sqrt{\alpha}$ in the $\alpha \rightarrow \infty$ limit, so that 
the product with the factor $\sqrt{\alpha}$ remains constant.  If such 
an asymptotic regime is observed at some large $\alpha$, then this is 
an indication that the solution process can be stopped.

\section{The short-time propagator}
\label{sec:four}

\begin{figure}[t]
\begin{center}
\leavevmode
\hbox{
\epsfxsize=8.5cm
\epsffile{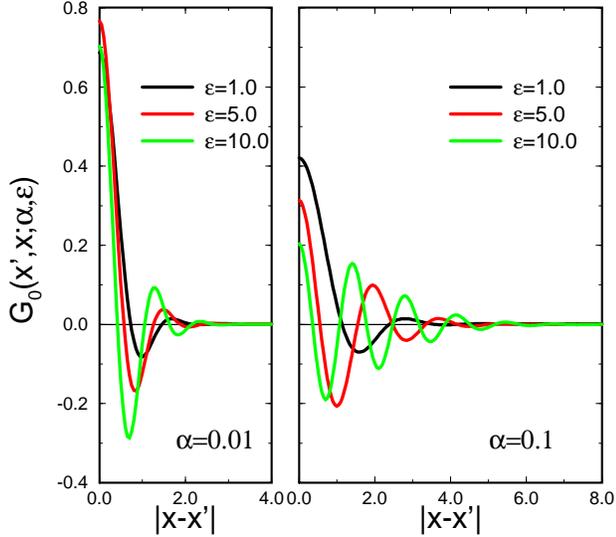}
}
\end{center}
\caption{
The free one-dimensional Gren's function (\ref{twenty}) for two 
values of the time interval $\alpha$ and three values of the
excitation energy $\varepsilon$.  The oscillations grow with 
both $\alpha$ and $\varepsilon$.
}
\label{fig3}
\end{figure}

Equation (\ref{five}) can be solved analytically in a handful of cases.  
In many-body problems, the equation becomes multi-dimensional and 
therefore requires stochastic methods of solution.  The central task
here is the calculation of the Green's function (GF) of the evolution
operator:
\begin{equation}
G({\bf X'},{\bf X};\alpha,\varepsilon) = 
\langle {\bf X'}| e^{-\alpha(\varepsilon-H)^2} |{\bf X}\rangle ,  
\label{fifteen}
\end{equation}
where ${\bf X} \equiv \{ x_i \}$ and ${\bf X'} \equiv \{ x'_i \}$ are 
two sets of the system's coordinates.  The number of coordinates is
$n = Nd$, $N$ being the number of particles and $d$ the dimensionality
of space.  I have also used the notation $\varepsilon$ to represent the 
sum $\omega+E_G$.  Let us find GF for the system of free distinguishable
particles with unit mass which is described by Hamiltonian
\begin{equation}
H_0 = \hat T \equiv -\frac{1}{2} \sum^n_{i=1} 
\frac{\partial^2}{\partial x^2_i} = -\frac{1}{2} \nabla^2_n ,  
\label{sixteen}
\end{equation}
where $\nabla_n$ is the $n$-dimensional gradient operator.  Introducing
the momentum states $|{\bf K}\rangle$ via
\begin{equation}
|{\bf X}\rangle = \int^{\infty}_{-\infty} \frac{d\,{\bf K}}{(2\pi)^n} 
e^{i{\bf KX}} |{\bf K}\rangle , 
\label{seventeen}
\end{equation}
and using $\langle {\bf K'} | {\bf K} \rangle = 
(2\pi)^n \delta({\bf K}-{\bf K'})$ one obtains from (\ref{fifteen})
\begin{equation}
G_0({\bf X'},{\bf X};\alpha,\varepsilon) = 
\int^{\infty}_{-\infty} \frac{d\,{\bf K}}{(2\pi)^n} 
e^{i{\bf K} ({\bf X}-{\bf X'})}
 e^{-\alpha(\varepsilon-\frac{{\bf K}^2}{2})^2} .  
\label{eighteen}
\end{equation}
The last expression satisfy the initial condition 
$G_0({\bf X'},{\bf X};\alpha=0) = \delta({\bf X}-{\bf X'})$.  
As expected from the translational invariance, GF is a function
of the difference $({\bf X}-{\bf X'})$.  The rotational invariance
of space also suggests that GF must be the function of the modulus
$|{\bf X}-{\bf X'}| \equiv R$ only.  This is indeed the case which
allows the reduction of Eq.~(\ref{eighteen}) to a one-dimensional
integral (technical details are not important for the purposes of 
this paper):
\begin{equation}
G_0(R;\alpha,\varepsilon) =
\frac{2\pi^{\frac{n}{2}}}{(2\pi)^n} 
\int^{\infty}_0 dk \, k^{n-1}
\frac{J_{\frac{n}{2}-1}(kR)}{(\frac{kR}{2})^{\frac{n}{2}-1}}
e^{-\alpha(\varepsilon-\frac{k^2}{2})^2} ,  
\label{nineteen}
\end{equation}
where $J_{\nu}(x)$ is the Bessel function.  In one dimension,
$n=1$, it follows from Eq.~(\ref{nineteen}) or directly from
Eq.~(\ref{eighteen}) that
\begin{equation}
G^{n=1}_0(x',x;\alpha,\varepsilon) = \int^{\infty}_{-\infty} 
\frac{dk}{2\pi} \cos[{k(x-x')}] \, 
e^{-\alpha(\varepsilon-\frac{k^2}{2})^2} .    
\label{twenty}
\end{equation}
The last equation is a convenient place to compare the properties
of the imaginary time Schr\"odinger equation (or diffusion equation)
with those of Eq.~(\ref{five}).  In the former case, the corresponding
expression for $G^{n=1}_0$ would have just $k$ in place of 
$(\varepsilon-\frac{k^2}{2})$.  Then, the integral would easily be 
performed leading to the positive-definite GF of the diffusion 
equation.  In the present case, despite the integral cannot be performed
analytically, it is easy to see that GF is {\em not} positive definite.
Indeed, take the limit of large $\alpha$.  The intergal is dominated
by the points $k = \pm \sqrt{2\varepsilon}$ and therefore as a function 
of $|x-x'|$ oscillates with period $(2\pi)/\sqrt{2\varepsilon}$.
This is the source of the negative regions and of the ``space''
sign problem.  Notice that the period is large for small $\omega$ 
and small for large $\omega$.  Thus the sign problem is expected to
be less severe in the physically interesting regions of low-energy
excitations.  This trend is encouraging and it is opposite to what 
happens in the time formulation, as discussed in Section~\ref{sec:two}.     
For the purposes of this paper it is more interesting to study
the case of small $\alpha$ and large distances $|x-x'|$.  A rigorous
analysis is quite involved, let me only mention that the amplitude of 
the GF decays exponentially in this limit:
\begin{equation}
|G^{n=1}_0(x',x;\alpha)| \leq 
C_1 \exp{\left( -C_2 |x-x'|^{4/3} \right) } ,    
\label{twentyone}
\end{equation}
where $C_1, C_2 > 0$ \cite{Fedoriuk}.  Figure~\ref{fig3} shows the results 
of numerical integration of Eq.~(\ref{twenty}) for several $\alpha$
and $\varepsilon$.  Notice that the number of oscillations increases
with the excitation energy.  $\alpha$ seems to affect the amplitude
of the oscillations rather than their period.  In conclusion of this 
part one should mention that the existence of the negative regions in 
the Green's function was shown to be a general property of differential 
operators with the order of spatial derivatives higher than 2 
\cite{Foulkes_two}.

Let me turn to the more difficult and more interesting case of
interacting quantum mechanical system.  The full Hamiltonian now
consists of the kinetic and potential parts, $H = \hat T + \hat V$.
The potential energy $\hat V$ may include single-body as well as
two-body terms.  As it is always done in Monte Carlo studies,
the long-time propagator (\ref{fifteen}) can be expressed as a 
multiple integral of the short-time propagator
\begin{eqnarray}
 G({\bf X'},{\bf X};\alpha) & =          & \nonumber \\
 \int d{\bf X}_1 \ldots     & d{\bf X}_L &
\; G({\bf X'},{\bf X}_L; \triangle \alpha) \ldots 
G({\bf X}_1,{\bf X}; \triangle \alpha)  .      
\label{twentytwo}
\end{eqnarray}
The number of additional time slices $L$ must be chosen large 
so that the new time interval $\triangle \alpha = \frac{\alpha}{L+1}$ 
gets small enough to allow accurate approximations of 
$G({\bf X}_{i+1},{\bf X}_i; \triangle \alpha)$.  The main difficulty
in constructing those approximations is the non-commutativity of the 
kinetic and potential operators $\hat T$ and $\hat V$.  This difficulty
is usually overcome by the Trotter-Suzuki decomposition of exponential
operators, see, e.g., \cite{deRaedt}.  However, in the present case one 
is faced with a completely new situation.  The exponent of the evolution
operator [the right hand side of Eq.~(\ref{five})] contains not only
sums but also {\em products} of non-commuting operators, namely 
$\hat T \hat V$ and $\hat V \hat T$.  Thus all the recipies have to be
devised anew.  In the first order in $\triangle \alpha$ it can be done
in the following way.  Expand the exponential in Eq.~(\ref{fifteen}):
\begin{eqnarray}
& G &({\bf X}_{i+1},{\bf X}_i;\triangle\alpha,\varepsilon) = 
\label{twentythree} \\ 
\langle &{\bf X}&_{i+1}| \left\{
1 -\triangle\alpha (\varepsilon -\hat T -\hat V)(\varepsilon -\hat T -\hat V)
+ o(\triangle\alpha) \right\}
|{\bf X}_i \rangle . \nonumber   
\end{eqnarray}
In the term linear in $\triangle \alpha$, the potential operator
in the first parantheses can be replaced with its value at the 
final configuration: $\hat V \rightarrow V({\bf X}_{i+1}) \equiv V_{i+1}$.
Similarly, the potential operator in the second parantheses can be 
replaced with its value at the initial configuration:
$\hat V \rightarrow V({\bf X}_i) \equiv V_i$.  Exponentiating back,
one obtains
\begin{eqnarray}
& G &({\bf X}_{i+1},{\bf X}_i;\triangle\alpha,\varepsilon)  
\nonumber \\ 
& \approx & \langle {\bf X}_{i+1}| e^{-\triangle\alpha 
(\varepsilon - V_{i+1} - \hat T)(\varepsilon - V_i - \hat T) }
|{\bf X}_i \rangle  \nonumber \\
& = & e^{\frac{\triangle\alpha}{4}(V_{i+1}-V_i)^2}
\langle {\bf X}_{i+1}| e^{-\triangle\alpha 
\left[ \varepsilon - \frac{1}{2}(V_{i+1}+V_i) - \hat T \right] ^2 } 
|{\bf X}_i \rangle 
\label{twentyfour}  \\
& = &  e^{\frac{\triangle\alpha}{4}(V_{i+1}-V_i)^2}
G_0 \left( {\bf X}_{i+1},{\bf X}_i;\triangle\alpha,
\varepsilon-\frac{1}{2}(V_{i+1}+V_i) \right) .
\nonumber   
\end{eqnarray}
The last line expresses the short-time propagator of the interacting
system via the short-time propagator of the non-interacting system
$G_0$ taken at a {\em shifted energy argument}.  This result together
with Eq.~(\ref{nineteen}) lead to the following remarkable conclusion:
{\em even in a many-body interacting system calculation of the 
short-time propagator requires knowledge of the potential energy
at the end points and only one one-dimensional integration.}
Moreover, since the integral in Eq.~(\ref{nineteen}) is a universal
function for given $n$ and $\alpha$, it can be easily tabulated for
all required values of $\varepsilon$ and $|{\bf X}_{i+1}-{\bf X}_i|$.

Of fundamental importance is the question whether the higher order
approximations could be derived.  Preliminary studies suggest a 
positive answer albeit with much more cumbersome expressions than
Eq.~(\ref{twentyfour}).  Continuing the expansion in 
Eq.~(\ref{twentythree}) further one finds in the next term already 
four factors $(\varepsilon - \hat T - \hat V)$.  Only two operators
$\hat V$ can be replaced with numbers immediately, the other two have
to be commuted through $\hat T$s.  Such a process generates multiple
additional terms that contain $(\nabla_n V)$s and $(\nabla^2_n V)$s.
I do not show specific details here but it is clear that the second
order approximation would be much harder to evaluate numerically
than the first order one, especially in the multi-dimensional
situation.  Obviously, a special study is needed to investigate
whether such a complexity is worth the gained accuracy and the 
consequent increase of the time step $\triangle \alpha$.

\section{Integration}
\label{sec:five}

With the short-time propagator known for given excitation energy 
$\omega$ and time step $\triangle \alpha$, the ``$\alpha$-dependent
spectral function'' is given by the multiple integral
\begin{eqnarray}
A(\alpha) & = (2\pi) &\sqrt{\frac{\alpha}{\pi}} 
\int^{\infty}_{-\infty} d{\bf X} d{\bf X}_1 \ldots  d{\bf X}_L d{\bf X'} 
\times  \nonumber \\ 
& \Phi({\bf X}') & \! G({\bf X'},{\bf X}_L;\triangle \alpha) \ldots 
G({\bf X}_1,{\bf X};\triangle \alpha) \, \Psi({\bf X})  ,      
\label{twentyfive}
\end{eqnarray}
where $\Psi$ and $\Phi$ are the initial and final states of the system.
(Recall that $\Psi$ and $\Phi$ may not be the same, this depends upon
the choice of the operators $O_1$ and $O_2$ studied.)  While trying 
to evaluate the above integral one faces an immediate difficulty of
infinite limits of integration.  It can be resolved by introducing
auxiliary {\em normalized} probability densities $P({\bf X})$ with
properties $P({\bf X}) > 0$ and $\int d{\bf X} P({\bf X}) = 1$.
Then the last equation is rewritten identically as follows
\begin{equation}
\frac{A(\alpha)}{2\pi}  = \! \sqrt{\frac{\alpha}{\pi}} \, 
\frac{
\int d{\bf X} d{\bf X}_1 \ldots d{\bf X'} \, W \,
P'({\bf X'}) \ldots P_1({\bf X}_1) P({\bf X})
}
{
\int d{\bf X} d{\bf X}_1 \ldots d{\bf X'} \:
P'({\bf X'}) \ldots P_1({\bf X}_1) P({\bf X})
}
\label{twentysix}
\end{equation}
\begin{equation} 
W \equiv \frac{
\Phi({\bf X}') G({\bf X'},{\bf X}_L;\triangle \alpha) \ldots 
G({\bf X}_1,{\bf X};\triangle \alpha) \, \Psi({\bf X}) 
}
{
P'({\bf X'}) P_L({\bf X}_L) \ldots P_1({\bf X}_1) P({\bf X})
} .      
\label{twentyseven}
\end{equation}
As a result of this transformation, the product of functions 
$P$ can serve as the weight function.  The explicit form 
of functions $P$ is completely at our disposal.  It should be
chosen to minimize the probability of large deviations of ${\bf X}$s.
The functions $P$ need not be all the same.  Each $P$ is absolutely 
arbitrary and independent of the other as long as the main conditions 
of positivity and normalization are fulfilled.  In short, functions 
$P$ serve to optimize stochastic integration of Eq.~(\ref{twentysix}).     
Note also that the absolute values of the Green's functions $|G|$
{\em cannot} be used as auxiliary probability densities because
their normalization is not known a priori.  [One may think that
one could get away from this obstacle by simply assuming some
normalization factor and then restoring it {\em afterwards} from
the overall normalization of the spectral function.  While this
can work in principle, this would require a very accurate determination 
of $A$ even at those energies one is not interested in.  Besides,
not every spectral function satisfies a normalization sum rule.]         

The actual measured quantity is the ratio $W$ from 
Eq.~(\ref{twentyseven}).  Since $G$s have negative regions their
product can be both positive and negative.  The sign alternation
of $W$ is going to be the main technical difficulty in implementing
the present idea in practice.

\section{Perturbed harmonic oscillator: Numerical solution}
\label{sec:six}

The results obtained in the last two sections will now be applied
to the harmonic oscillator example discussed above.  Two issues 
are going to be addressed specifically: (i) the accuracy of the
first-order approximation (\ref{twentyfour}) and (ii) stability
of the integration in (\ref{twentysix})-(\ref{twentyseven}).

\begin{figure}[t]
\begin{center}
\leavevmode
\hbox{
\epsfxsize=8.5cm
\epsffile{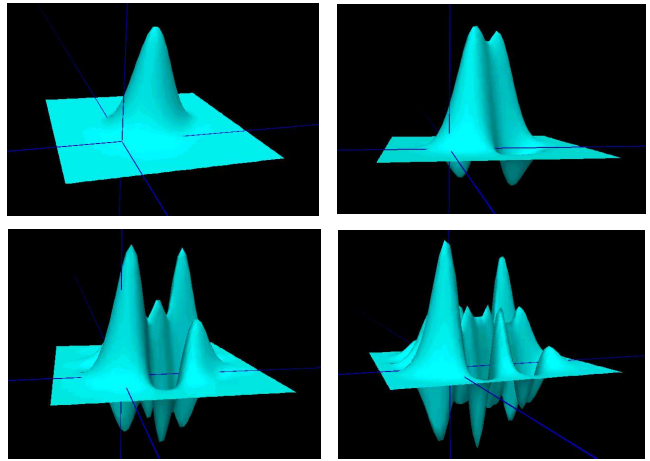}
}
\end{center}
\caption{
The Green function (\ref{twentyeight}) for $\alpha = 2.0 \, \Omega^{-2}$ 
and $\omega=-2.75 \, \Omega$ (top left), $\omega = -1.75 \, \Omega$ 
(top right), $\omega = -0.25 \, \Omega$ (bottom left), and 
$\omega = 1.25 \, \Omega$ (bottom right).  The number of nodes increases 
with excitation energy.  The two horizontal axes are $x$ and $x'$. 
}
\label{fig4}
\end{figure}

The exact oscillator propagator for {\em any} time interval
$\alpha$ is given by
\begin{eqnarray}
G(x',x;\alpha,\omega) & = & \frac{1}{\sqrt{\pi}} 
e^{-\frac{1}{2}(x'-\sqrt{2}g)^2} e^{-\frac{1}{2}(x-\sqrt{2}g)^2}
\times \label{twentyeight}  \\
\sum^{\infty}_{m=0} \frac{1}{2^m m!} 
H_m(& x' & - \sqrt{2}g) H_m(x-\sqrt{2}g)
e^{-\alpha(\omega+g^2\Omega-\Omega m)^2} . 
\nonumber 
\end{eqnarray}
This function is shown in Figure~\ref{fig4} for 
$\alpha = 2.0 \, \Omega^{-2}$ and several excitation energies 
$\omega$.  Notice how the number of sign changes increases 
with $\omega$.  An analogy seems to exist between this fact and
the increase in the number of nodes of the eigenstates of the 
time-independent Schr\"odinger equation.  

One can use the exact solution (\ref{twentyeight}) to check the accuracy
of the approximate representation (\ref{twentyfour}).  Due to the 
one-dimensional character of the problem the check can be very 
effectively done as follows.  The approximate short-time propagator
is computed for some small time step $\triangle \alpha$.  
Then the propagator corresponding to the double time 
$G(2\triangle \alpha)$ is found by convolving two $G(\triangle \alpha)$s.
If $G(x',x;\alpha)$s are represented by matrices with matrix elements
evaluated on a uniform mesh $(x'_i,x_j)$ then the convolution is
equivalent to {\em squaring} of matrix $G(x'_i,x_j;\triangle \alpha)
\equiv G_{ij}(\triangle \alpha)$:
\begin{equation}
\hat G_{ij}(2\triangle\alpha) = \triangle x \, 
\hat G^2_{ij}(\triangle\alpha) , 
\label{twentynine} 
\end{equation}
where $\triangle x$ is the step of the coordinate mesh.  Repeating 
the squaring $n$ times one arrives at the propagator corresponding to
the large time interval $2^n \triangle \alpha$, which is compared with 
the exact result.  In this manner it was established that the rule 
(\ref{twentyfour}) is not as accurate as the one {\em without}
the exponential factor 
$\exp [\frac{\triangle \alpha}{4}(V_{i+1}-V_i)^2 ]$.  For instance,
at $\triangle \alpha = 0.05 \, \Omega^{-2}$ the expression without 
the factor remains within 10\% from the exact solution even after
9 squaring which corresponds to 512 time steps or total time 
$\alpha = 25.6 \, \Omega^{-2}$, in a wide range of energies 
$-2.75 < \omega/\Omega < 2.75$.  At the same time, the formula
with the exponential factor demonstrated deviations $\sim 100\%$ already
at 7 squaring, i.e., at times $\alpha \approx 6.4 \, \Omega^{-2}$.   
At larger $\triangle \alpha \approx 0.2-0.4 \, \Omega^{-2}$, both
versions were bad, Eq.~(\ref{twentyfour}) overestimating and the
one without the factor underestimating the correct values of the
long-time propagator.  The best results were obtained with the factor
but with the coefficient in the exponent being $\approx 0.10$ instead 
of 0.25 as in Eq.~(\ref{twentyfour}). 

It is not the purpose of the paper to perform a detailed examination
of this issue.  Given the complex character of the new evolution
operator, a complete analysis could be quite involved.  A possible 
reason for the described inconsistency may be a bigger than expected 
role of the higher-order terms neglected in deriving approximation 
(\ref{twentyfour}).  A better understanding of this role requires 
explicit derivation of the second and possibly third-order
approximations and their thorough comparison.  This is a difficult 
task that warrants a separate study. 

\begin{figure}[t]
\begin{center}
\leavevmode
\hbox{
\epsfxsize=8.5cm
\epsffile{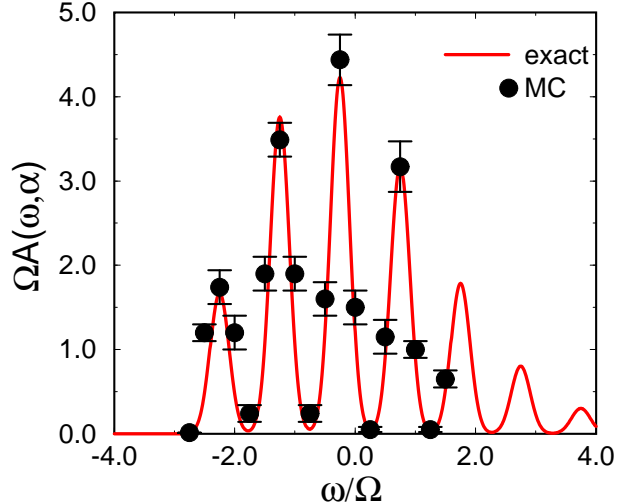}
}
\end{center}
\caption{
Results of Monte Carlo integration \, of 
Eqs.~(\ref{twentysix})-(\ref{twentyseven}) for the disturbed
harmonic oscillator.  The time step 
$\triangle \alpha = 2.0 \, \Omega^{-2}$ and the exact propagator
(\ref{twentyeight}) are used, see also Figure~\ref{fig4}.  The full 
time interval is $\alpha = 20.0 \, \Omega^{-2}$.  The total number 
of integrations is 11.  The number of measurements is $10^6$. 
}
\label{fig5}
\end{figure}

Let me now present the results of numerical integration of 
Eqs.~(\ref{twentysix})-(\ref{twentyseven}).  I have used gaussian
functions $P(x)$ and the simplest ``brute force'' method of 
integration when {\em uncorrelated} sequence of configurations 
$\{ x \}$ is generated and the quantity $W$ measured and
accumulated in a straightforward manner.  A crucial parameter 
in the process is the number of time intervals or the total 
number of integrations.  As long as the number of integrations
is less than about 20 the sign problem does not manifest itself
and the simulations produce accurate results with
controllable error bars.  An example is shown in Figure~\ref{fig5}
where the exact propagator (\ref{twentyeight}) is used.
The multipeak structure of the spectral function is clearly resolved.
Similar encouraging results have been obtained for other time steps
and number of steps, for instance 
$\triangle \alpha = 0.8 \, \Omega^{-2}, N=16$; 
$\triangle \alpha = 1.0 \, \Omega^{-2}, N=20$; and so on.
However, when the number of integrations exceeds 20, the sign
problem sets in and the simulations cannot be completed. 

\begin{figure}[t]
\begin{center}
\leavevmode
\hbox{
\epsfxsize=8.5cm
\epsffile{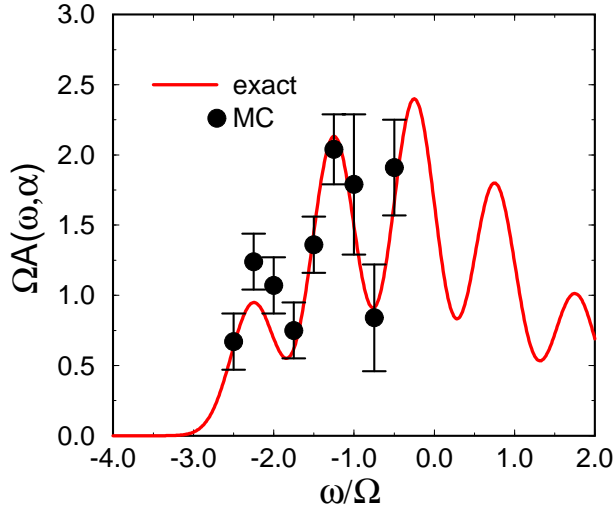}
}
\end{center}
\caption{
Results of Monte Carlo integration \, of 
Eqs.~(\ref{twentysix})-(\ref{twentyseven}) for the disturbed harmonic 
oscillator.  The time step $\triangle \alpha = 0.4 \, \Omega^{-2}$ and 
the approximate propagator (\ref{twentyfour}) are used.  The full 
time interval is $\alpha = 6.4 \, \Omega^{-2}$.  The total number of 
integrations is 17. (This is the reason for larger error bars as
compared with Figure~\ref{fig5}.)  The number of measurements is 
$10^7$. 
}
\label{fig6}
\end{figure}

The results of Monte Carlo integration with the approximate 
propagator (\ref{twentyfour}) are shown in Figure~\ref{fig6}.  
In this example, the numerical coefficient in the exponent of 
the exponential factor is taken to be 0.10.  The number of 
time steps is 16 and the total ``time'' is 
$\alpha = 6.4 \, \Omega^{-2}$.  Error bars have been estimated 
from the results of 5 successive runs of $10^7$ measurements.
The agreement between the Monte Carlo data and the exact
spectral function is good, given the simplicity of the numerical
method used.  In this respect several comments are appropriate.
First of all, the perturbed harmonic oscillator is not a simple
system as far as its spectral function is concerned.  The latter
has a multi-peak structure and is a hard task for any method.
It is known that the inability to effectively resolve multiple 
peaks is the main shortcoming of the mainstream reconstruction
methods.  The very fact that the present method {\em is} able
to resolve multiple peaks is very encouraging.  This is a consequence
of being a direct method: there is no difference between the
smooth and the peaked function as long as the value for the
given energy point is computed independently and on the same 
footing with other points.  Secondly, it has not been the purpose 
of this paper to provide a full-scale numerical analysis of
this example.  The main interest lies in multi-dimensional
interacting systems which nevertheless may have simpler structure
functions.  There is not much sence in perfecting the solution
when there is no guarantee that the tricks developed will work
in the physically more interesting cases.  The results given
above serve primarily as an illustration to the method and a 
demonstration of its viability.  Thirdly, and perhaps most 
importantly, the ``brute force'' method of integration
is {\em not} the right way to solve the equation (\ref{five}).
The incredible progress of Monte Carlo methods in the past 
decades has been largely due to the ``smart'' ways of performing
multiple integrations such as the Metropolis algorithm in
statistical applications or the importance sampling in 
fermionic applications.  In relation to the present paper,
the recent exact fermionic algorithm of Kalos and Pederiva 
\cite{Kalos} looks particularly interesting.  They developed a 
Diffusion Monte Carlo algorithm for finding the ground state of the 
many-body fermionic Schr\"odinger equation with no approximation.
Their method can handle wave functions with positive and negative 
regions, so it effectively solves the fermionic sign problem.
Precisely an algorithm of this kind is required for solving
equation (\ref{five}).  It is too early to judge whether these 
ideas could be directly applied here but this is definitely one
of possible directions to search for an effective method of
integration.

\section{Conclusions}
\label{sec:seven}

This paper describes a novel approach to direct stochastic 
calculation of dynamic correlation functions of quantum mechanical 
systems.  It was inspired by the difficulties of the traditional 
reconstruction methods originating from the absence of a
mathematically rigorous formulation of the analytical continuation
procedure.  I believe that at present the quest for an effective 
way of computing the dynamic correlators directly in real 
time/frequency domain is fully justified, and it should continue
until any of the two approaches results in an effective and 
satisfactory numerical algorithm.  

Let me summarize how the present paper contributes to this process.
It has been proposed to make the spectral function $A(\omega)$ rather 
than the time-dependent correlator $K(t)$, the main object of the 
computational effort.  This is advantageous because it directly leads to 
experimentally accessible properties and avoids the need for additional 
procedures such as Fourier transformations.  The calculation of the 
spectral function is made possible by replacing the operator 
delta-function with the limit of a gaussian and by interpreting the 
latter as a long-time solution of the evolution equation 
(\ref{five}).  Thus the process of finding $A(\omega)$ is 
formulated as an initial value problem for a fourth-order differential 
operator.  Such a problem is {\em mathematically well-defined} and has 
a unique solution.  One ``only'' requires to solve the equation in the 
limit of infinite time $\alpha$.  The solutions can be classified with 
respect to their limit behavior.  Some will remain constant which will 
lead to infinite values of the spectral function (apparently at discrete 
energies only).  Other solutions should decay as 
$\propto \frac{1}{\sqrt{\alpha}}$ and lead to finite $A(\omega)$.
Third class comprises solutions decaying faster than that (probably 
exponentially) and resulting in vanishing $A(\omega)$.  In the end,
taking a scalar product with a function of interest is required.  
The whole sequence should be repeated for all relevant excitation 
energies.  The new formulation has been shown to be equivalent
to the standard theory and definitions of the spectral function.
It has also been generalized to finite temperatures.  Another
positive feature of the method is its universality.  It imposes
absolutely no restrictions on the nature of the operators studied.
They could be bosonic or fermionic, one-particle, two-particle,
or even more complex, current, density, spin, and so on.  Also the 
spectral functions are not required to possess any special properties
such as non-negativity, additional symmetries, and sum rules.
In short, the theory is simple, mathematically well-defined, and
universal. 

The practical usefulness of the proposed method will be very much 
dependent on the availability of stochastic methods that could solve
equation (\ref{five}) in multi-dimensional cases.  Since the propagator 
necessarily has negative regions, handling of sign-alternating 
functions is required.  An example considered in Section~\ref{sec:six} 
demonstrates that this is achievable.  In broader terms, the situation 
bears strong similarities with Diffusion Monte Carlo studies of systems 
of interacting fermions, where a significant progress was reported 
\cite{Kalos,Anderson}.  Additional work is needed to develop similar
approaches for equation (\ref{five}).  If successful, the effort will
bring direct ways of first-principle Monte Carlo studies of
dynamic characteristics of quantum mechanical systems.

\acknowledgements

I am grateful to Matthew Foulkes for illuminating discussions on the
subject.


\begin{references}

\bibitem{Foulkes}
W.\,M.\,C.\,Foulkes, L.\,Mitas, R.\,J.\,Needs, and G.\,Rajagopal,
Rev.\,Mod.,Phys. {\bf 73}, 33 (2001).

\bibitem{Ceperley}
D.\,M.\,Ceperley,
Rev.\,Mod.\,Phys. {\bf 67}, 279 (1995).

\bibitem{Krauth}
W.\,Krauth,
Phys.\,Rev.\,Lett. {\bf 77}, 3695 (1996).

\bibitem{Magro}
W.\,R.\,Magro, D.\,M.\,Ceperley, C.\,Pierleoni, and B.\,Bernu,
Phys.\,Rev.\,Lett. {\bf 76}, 1240 (1996).

\bibitem{Leung}
W.-K.\,Leung, R.\,J.\,Needs, G.\,Rajagopal, S.\,Itoh, and S.\,Ihara,
Phys.\,Rev.\,Lett. {\bf 83}, 2351 (1999).

\bibitem{Bonisegni}
M.\,Bonisegni and D.\,M.Ceperley,
Phys.\,Rev.\,Lett. {\bf 74}, 2288 (1995).

\bibitem{Evertz}
H.\,G.\,Evertz, G.\,Lana, and M.\,Marcu,
Phys.\,Rev.\,Lett. {\bf 70}, 875 (1993).

\bibitem{Blankenbecler}
R.\,Blankenbecler, D.\,J.\,Scalapino, and R.\,L.\,Sugar,
Phys.\,Rev.\,D {\bf 24}, 2278 (1981).

\bibitem{Kornilovitch}
P.\,E.\,Kornilovitch,
Phys.\,Rev.\,Lett. {\bf 81}, 5382 (1998).

\bibitem{Prokofev}
N.\,V.\,Prokof'ev and B.\,V.\,Svistunov,
Phys.\,Rev.\,Lett. {\bf 81}, 2514 (1998).

\bibitem{Jarrell}
M.\,Jarrell and J.\,E.\,Gubernatis,
Phys.\,Rep. {\bf 269}, 133 (1996), 
and references therein.

\bibitem{Vidberg}
H.\,J.\,Vidberg and J.\,W.\,Serene,
J.\,Low\,Temp.\,Phys. {\bf 29}, 179 (1977).

\bibitem{Creffield}
C.\,E.\,Creffield, E.\,G.\,Klepfish, E.\,R.\,Pike, and S.\,Sarkar,
Phys.\,Rev.\,Lett. {\bf 75}, 517 (1995).

\bibitem{Mishchenko}
A.\,S.\,Mishchenko, N.\,V.\,Prokof'ev, A.\,Sakamoto, and 
B.\,V.\,Svistunov,
Phys.\,Rev.\,B {\bf 62}, 6317 (2000).

\bibitem{Mak}
C.\,H.\,Mak and R.\,Egger,
J.\,Chem.\,Phys. {\bf 110}, 12 (1999).

\bibitem{Egger}
C.\,H.\,Mak, R.\,Egger, and H.\,Weber-Gottschick,
Phys.\,Rev.\,Lett. {\bf 81}, 4533 (1998).

\bibitem{Matthew}
W.\,M.\,C.\,Foulkes 
(private communication).

\bibitem{Mahan}
G.\,D.\,Mahan,
{\em Many-Particle Physics}, 2nd ed. (Plenum Press, 1990).

\bibitem{Fedoriuk}
M.\,V.\,Fedoriuk,
{\em Asymptotic Analysis of Integrals and Series}, in Russian
(Nauka, Moscow, 1987, Library of Congress Control Number: 88162273).  

\bibitem{Holstein}
T.\,Holstein,
Ann.\,Phys. {\bf 8}, 325 (1959).

\bibitem{Foulkes_two}
W.\,M.\,C.\,Foulkes and M.\,Schluter,
Phys.\,Rev.\,B {\bf 42}, 11505 (1990). 

\bibitem{deRaedt}
H.\,deRaedt and B.\,deRaedt,
Phys.\,Rev.\,A {\bf 28}, 3575 (1983).  

\bibitem{Kalos}
M.\,H.\,Kalos and F.\,Pederiva,
Physica A {\bf 279}, 236 (2000);
Phys.\,Rev.\,Lett. {\bf 85}, 3547 (2000). 

\bibitem{Anderson}
J.\,B.\,Anderson, C.\,A.\,Traynor, and B.\,M.\,Boghosian,
J.\,Chem.\,Phys. {\bf 95}, 7418 (1991).  

\end{references}
\end{document}